\documentclass[aip,jcp,10pt,twocolumn,superscriptaddress,nofootinbib,reprint]{revtex4-1}

\usepackage{bm}
\usepackage[english]{babel}
\usepackage{graphicx}
\usepackage{hyperref}
\usepackage[utf8]{inputenc}
\usepackage{mathtools}
\usepackage[expansion=true,protrusion=true]{microtype}
\usepackage{siunitx}
\usepackage{textcomp}
\usepackage{textgreek}
\usepackage[usenames,dvipsnames]{xcolor}

\DeclareSIUnit\Molar{M}
\hypersetup{hidelinks}
\usepackage{cleveref}

\renewcommand{\eqref}[1]{(Eq.\,\ref{#1})}
\crefformat{equation}{(Eq.\,#2#1#3)}

\newcommand{\lD}{\lambda_{D}}
\newcommand{\kT}{k_{B}T}

\let\vec\bm

\usepackage{natbib}

\begin{document}

\title{Selective Trapping of DNA using Glass Microcapillaries}

\author{Georg Rempfer}
\email{georg.rempfer@icp.uni-stuttgart.de}
\author{Sascha Ehrhardt}
\affiliation{Institute for Computational Physics (ICP), University of Stuttgart, Allmandring 3, 70569 Stuttgart, Germany}
\author{Nadanai Laohakunakorn}
\altaffiliation{Current address: EPFL STI IBI-STI LBNC, BM 2137 Station 17, CH-1015 Lausanne, Switzerland}
\affiliation{Cavendish Laboratory, University of Cambridge, Cambridge CB3 0HE, United Kingdom}
\author{Gary B. Davies}
\affiliation{Institute for Computational Physics (ICP), University of Stuttgart, Allmandring 3, 70569 Stuttgart, Germany}
\author{Ulrich F. Keyser}
\affiliation{Cavendish Laboratory, University of Cambridge, Cambridge CB3 0HE, United Kingdom}
\author{Christian Holm}
\affiliation{Institute for Computational Physics (ICP), University of Stuttgart, Allmandring 3, 70569 Stuttgart, Germany}
\author{Joost de Graaf}
\altaffiliation{Current address: School of Physics and Astronomy, University of Edinburgh, Scotland, Edinburgh EH9 3JL, United Kingdom.}
\affiliation{Institute for Computational Physics (ICP), University of Stuttgart, Allmandring 3, 70569 Stuttgart, Germany}

\date{\today}

\begin{abstract}

We show experimentally that a cheap glass microcapillary can accumulate \mbox{\textlambda-phage} DNA at its tip and deliver the DNA into the capillary using a combination of electro-osmotic flow, pressure-driven flow, and electrophoresis.
We develop an efficient simulation model for this phenomenon based on the electrokinetic equations and the finite-element method. Using our model, we explore the large parameter space of the trapping mechanism by varying the salt concentration, the capillary surface charge, the applied voltage, the pressure difference, and the mobility of the analyte molecules. Our simulation results show that this system can be tuned to capture a wide range of analyte molecules, such as DNA or proteins, based on their electrophoretic mobility. 
Our method for separation and pre-concentration of analytes has implications for the development of low-cost lab-on-a-chip devices.

\end{abstract}

\maketitle

Analytic biochemistry possesses a large variety of methods to purify and pre-concentrate macroscopic amounts of analyte molecules for further processing. These methods are based on precipitation, centrifugation, gel electrophoresis, chromatography, and affinity techniques, to name just a few.~\cite{ninfa_fundamental_2009, cuatrecasas_protein_1970} Lab-on-a-chip devices, on the other hand, have only nano- to micro-liters of solution at their disposal, and the resulting samples need to be processed on a similarly small scale, which poses a much greater challenge. Lab-on-a-chip devices have applications in, \emph{e.g.}, the low-cost and large-scale screening of DNA.~\cite{figeys_lab-on-a-chip_2000} In the last decade, a number of methods based on microfluidics and electrokinetic effects have been proposed to purify samples on the microscale. 

Asbury~\emph{et al.}\cite{asbury_trapping_1998, asbury_trapping_2002}\ constructed a device that traps DNA at the edges of thin gold electrodes using dielectrophoretic forces, creating a region of high DNA concentration. Dielectrophoresis involves the polarization of an object, such as DNA and its double layer, due to an electric field gradient, which causes the polarized object to move. By switching from the inhomogeneous trapping electric field to a homogeneous electric field, Asbury~\emph{et al.}\cite{asbury_trapping_1998, asbury_trapping_2002} transported DNA from one edge to the other through ordinary electrophoresis.

\citet{wang_million-fold_2005} created a microfluidic silica channel, filled with electrolytic analyte solution, connected by a T-junction to a smaller nanofluidic channel. An electric field directed along the analyte channel drives molecules near the T-junction through a combination of electrophoresis and electro-osmotic flow (EOF). Biasing the nanochannel with a negative voltage against the analyte channel creates an electrostatic barrier for negatively charged analyte molecules that concentrates them upstream of the T-junction. 

Following similar principles,~\citet{huang_digitally_1999} developed a device that transports different analytes through a microchannel by advection using pressure-driven flow. They created an electric field with increasing strength by placing electrodes along the channel. The increasing electrophoretic force stalls different molecules at different points along the channel based on their electrophoretic mobility.~\citet{stein_electrokinetic_2010} showed that this concentration effect also happens at the entrance of a nanopore, without the need for the additional electrodes or channels used by Huang~\emph{et al.}~\cite{huang_digitally_1999}

Finally, \citet{ying_programmable_2002, ying_frequency_2004} were able to achieve this concentration effect using cheap glass capillaries. They filled a tapered glass capillary of diameter~\SI{100}{\nano\m} with DNA and applied an outwardly-directed electric field. Through a combination of electro-osmotic, electrophoretic, and dielectrophoretic effects, they were able to concentrate DNA at the capillary tip. In addition to the accumulation of DNA,~\citet{ying_programmable_2002, ying_frequency_2004} showed that reversing the electric field delivers the concentrated DNA sample out of the capillary.

In this paper, we show, by experiment, how to concentrate DNA using glass microcapillaries similar to those used by \citet{ying_programmable_2002, ying_frequency_2004} However, in contrast to their mechanism based on dielectrophoresis, we use capillaries with orifice diameters of several micrometers in which dielectrophoresis plays no role.~\cite{ying_frequency_2004} Instead, we use a mechanism based on a combination of pressure-driven flow, EOF, and electrophoresis to control the DNA accumulation in a fashion more similar to the technique of Huang~\emph{et al.,}~\cite{huang_digitally_1999} but relying on cheap glass microcapillaries instead of comparatively elaborate lithography and etching techniques.

Using finite-element simulations of the electrokinetic equations,~\cite{white_ion_2008, laohakunakorn_electroosmotic_2015, laohakunakorn_electroosmotic_2015-1, lan_nanoparticle_2011, german_controlling_2013, lan_effect_2014, laohakunakorn_dna_2013, mao_hydrodynamic_2013} we determine the electrophoretic mobility of trapped particles depending on the applied voltage and pressure bias, as well as the salt concentration and the pore surface charge. Our results demonstrate that it is possible to maintain a high degree of control over the analyte accumulation despite the more complicated geometries of glass micro-capillaries. These results could pave the way for cheap microfluidics device capable of sequentially purifying, pre-concentrating, and delivering samples of different analyte molecules from a mixed solution.

\section{Experimental Setup and Methods}
\label{sec:experiment}

\begin{figure}
\includegraphics[height=0.7\columnwidth]{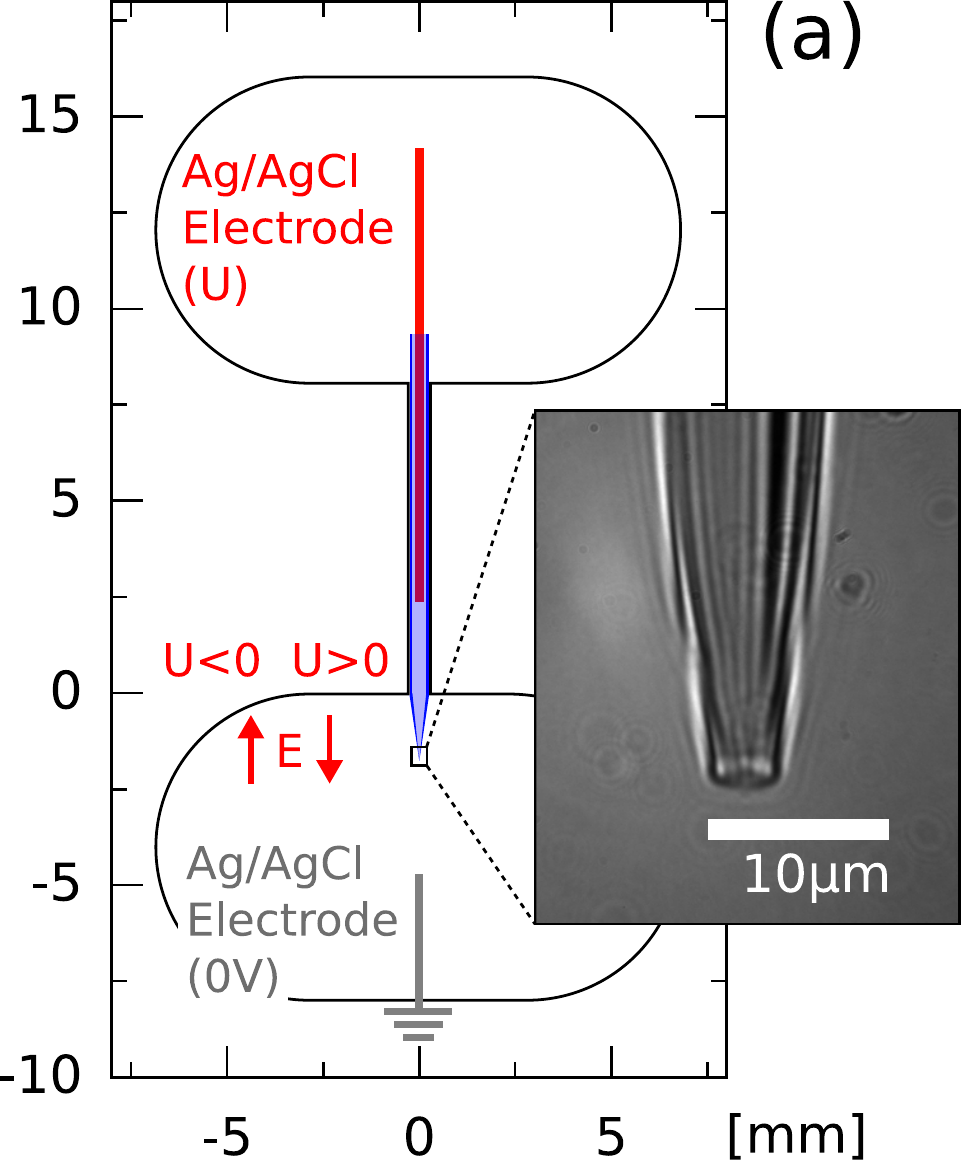}
\includegraphics[height=0.7\columnwidth]{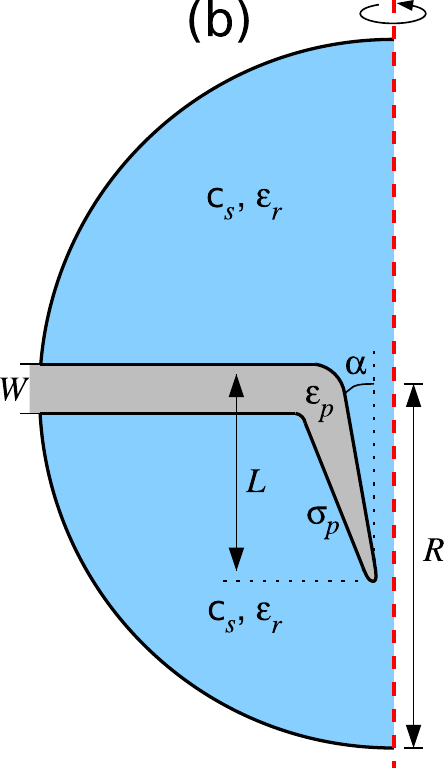}
\caption{(a) Sketch of the experimental setup consisting of two millimeter-sized reservoirs, connected by the glass capillary (blue). Also depicted is the positioning of the electrodes, with the working electrode (red) reaching into the back of the capillary and the reference electrode (gray) being located in the reservoir connected to the tapered capillary tip at a distance of several millimeters from the capillary orifice. The inset shows a light micrograph image of the tapered tip of the capillary with an orifice of diameter \SI{4}{\micro\m}.\\
(b) Sketch of the conical capillary geometry used in our numerical calculations. The domain is rotationally symmetric around the dashed red axis. There are two hemispherical reservoirs (blue), separated by a circular barrier of radius $R = 2.5 L$ and width $W = 0.15 L$ (grey), from which a conical pore extrudes of length $L = \SI{20}{\micro\m}$ and inner opening angle $\alpha = \ang{5}$. The pore and saline solution (bulk salt concentration $c_{s}$) are dielectrics with relative permittivity $\varepsilon_{\text{p}} = 4.2$ and $\varepsilon_{\text{r}} = 78.54$, respectively. The origin of the coordinate system is located at the narrowest part of the capillary (the orifice), with a radius $r_{\text{m}} = \SI{0.5}{\micro\m}$. The capillary carries a surface charge density $\sigma_{\text{p}}$, while the barrier is uncharged.\\}
\label{fig:system}
\end{figure}

We use two Ag/AgCl electrodes to apply a voltage between the back end of a glass capillary and the reservoir in contact with the capillary's tapered tip as shown in Fig.\,\ref{fig:system}a. The electrode located in the reservoir at a distance of several millimeters from the capillary orifice serves as the ground electrode providing the reference potential. In the following, we will refer to the voltage of the working electrode inserted into the back of the capillary as the voltage applied to the system. Positive voltages correspond to an electric field directed from the back of the capillary, through the tapered capillary tip and the capillary orifice, into the lower reservoir. Negative voltages correspond to an electric field in the opposite direction, from the lower reservoir, through the capillary orifice and tip, to the electrode in the back of the capillary. Different filling heights of the two reservoirs connected by the capillary allow us to establish a pressure difference between the two ends of the capillary.\\\indent
The reservoir surrounding the capillary tip contains \mbox{\textlambda-phage} DNA. Translocating DNA molecules through the capillary orifice modulates the ionic current between the two electrodes. We monitor this current and use its modulations to trigger the capturing of images using a fluorescence microscope. The fluorescence microscope, whose setup has previously been discussed~\cite{thacker_studying_2012} observes a small volume around the tapered end of the glass microcapillary.\\

\begin{footnotesize}
The microcapillary was produced by pulling apart borosilicate glass capillaries (Hilgenberg, inner diameter \SI{0.376}{\milli\m}, outer diameter \SI{0.5}{\milli\m}) using a programmable infrared laser puller (\mbox{P-2000}, Sutter Instruments) with the following protocol:\\
\texttt{Heat=167, Filament=2, Velocity=10, Delay=130, Pull=80,}\\
repeated once followed by\\
\texttt{Heat=220, Filament=0, Velocity=10, Delay=128, Pull=95,}\\
with the whole program looping twice. This produced tapered microcapillaries with orifice diameters ranging from 3 to \SI{10}{\micro\m}. The capillary is subsequently assembled into a polydimethylsilane (PDMS) matrix and mounted on a glass slide.\\\indent
Both reservoirs and the capillary are filled with an aqueous KCl solution (5 to \SI{1000}{\milli\Molar}) containing \mbox{\textlambda-phage} DNA (\SI{48}{\kilo bp}, New England Biolabs) and the intercalating cationic dye \mbox{POPO-3} iodide (Invitrogen) as well as \SI{10}{\milli\Molar} of $1\times$ Tris-EDTA buffer (Sigma Aldrich). \\\indent
The optimal dye-to-DNA ratio was determined by fluorimetry measurements to be 1:20 dye molecules to DNA base pairs, with a final DNA concentration of \SI{5}{\micro\Molar\,bp}, or roughly \SI{3.3}{\micro\g\,\milli\liter^{-1}}. The optimal dye-to-DNA ratio is dependent on salt concentration, and increases to around 1:6 at \SI{1}{\Molar} KCl.\\\indent
Photobleaching was observed, with a rate increasing as a function of laser intensity; at the highest laser powers used, complete photobleaching within the field of view occurred over a timescale of a few seconds. It is possible to reduce photobleaching by adding an antioxidant such as \mbox{\textbeta-mercaptoethanol}; however, we can mitigate these effects to an acceptable level by ensuring that the DNA is in continuous flow.

\end{footnotesize}

\section{Modeling and Simulation Methods}
\label{sec:simulations}

In this section we provide the details of our numerical modeling of the motion of DNA through a microcapillary. First, we discuss the principles that govern the motion of DNA, next we describe the means by which we determine the EOF and the electric field in the capillary geometry, and we conclude with a discussion of our numerical methods.

\subsection{DNA Motion}
\label{sec:dna_motion}

DNA in aqueous solution is strongly negatively charged and forms a positive double layer with an excess of $\text{K}^+$ ions and a characteristic length scale for its thickness, given by the Debye-length, and ranging from \SIrange{10}{0.3}{\nano\m} for typical salt concentrations $c_s$ of \SIrange{1}{1000}{\milli\Molar}. The Debye-length is given by $\lD = \sqrt{\varepsilon_0\varepsilon_\text{r}\kT / (2 N_A e^2 c_s)}$, where $\varepsilon_{\text{r}} = 78.54$ is the relative permittivity of water at temperature $T = \SI{298.15}{\kelvin}$, $\varepsilon_0$ is the vacuum permittivity, $k_B$ is Boltzmann's constant, $e$ is the elementary charge, and $N_A$ is Avogadro's constant.

In the presence of an electric field, the negatively charged DNA molecule and $\text{Cl}^-$ ions move against the local electric field, while most of the positively charged $\text{K}^+$ ions surrounding the DNA move in the direction of the electric field in a process called electrophoresis. Some $\text{K}^+$ ions are strongly bound to and co-moving with the DNA molecule, effectively reducing its charge.~\cite{manning_limiting_1981}

In addition to this relative motion of DNA and fluid, the DNA also advects with the local fluid velocity $\vec u$. The fluid velocity several Debye-lengths from the charged DNA molecule remains unaffected by the DNA molecule's presence.

The velocity of the DNA molecule $\vec v$ relative to the surrounding bulk fluid flow $\vec u$ scales linearly with the local electric field $\vec E$ for sufficiently small electric field magnitudes, which allows us to define an electrophoretic mobility $\mu$ for the DNA molecule.~\cite{huckel1924kataphorese, smoluchowski_contribution, obrien_electrophoretic_1978, stellwagen_free_1997} This mobility depends on the intricate coupling of ion motion, hydrodynamics, and electrostatics. The mobility decreases with increasing salt concentration and increases with increasing length of the DNA, but reaches a plateau at 400 DNA bases or more.~\cite{allison_length_2001} For a given mobility, the velocity of a DNA molecule is therefore:
\begin{equation}
\label{eq:point_particle}
\vec v = \vec u + \mu \vec E .
\end{equation}
In the experiments described in~\Cref{sec:experiment}, the fluid velocity is given by a superposition of the EOF and pressure-driven flow, which we discuss next.

\subsection{The Electrokinetic Equations}
\label{sec:ekin_eq}

In a cylindrical capillary with no tapering, the pressure-driven flow assumes the typical parabolic Poiseuille flow profile, while EOF leads to a plug flow profile with a constant flow velocity everywhere except in the double layers of the capillary walls and DNA molecules, where considerable shear allows the fluid to fulfill the no-slip boundary condition. The flow profiles in our conical capillaries deviate from this idealized situation due to pressure buildup in the axial direction.

To evaluate the flow $\vec u$ due to EOF and pressure differences between the reservoirs, as well as the electric field $\vec E$, we need to solve the coupled system of electrokinetic equations. We describe the electric field through the electrostatic potential $\Phi$ using Poisson's equation
\begin{equation}
\label{eq:ekin_electrostatics}
\nabla \cdot (\varepsilon_0\varepsilon_\text{r} \nabla \Phi) = -\varrho = -e(c_+ - c_-) .
\end{equation}
Here, the charge density $\varrho$ is given in terms of the concentrations $c_\pm$ of $\text{K}^+$ and $\text{Cl}^-$ ions, respectively, and the externally applied field enters through boundary conditions for the electric potential $\Phi$.

We model the ionic concentrations using a stationary diffusion-advection and continuity equation for each of the two ionic species:
\begin{equation}
\label{eq:ekin_ions}
\nabla \cdot (\underbrace{-D_\pm \nabla c_\pm - \mu_\pm z_\pm e c_\pm \nabla \Phi}_{\vec j_\pm^\text{diff}} + \underbrace{c_\pm \vec{u}}_{\vec j_\pm^\text{adv}}) = 0 .
\end{equation}
The ions' diffusion coefficients $D_{\pm} = \SI{2e-9}{\m^2\per\s}$ are related to their respective mobilities $\mu_\pm$ through the Einstein-Smoluchowski relation $D_\pm/\mu_\pm = \kT$.
The second and third term model the flux of ions relative to the fluid due to an electric field $\vec E = -\nabla \Phi$, and the flux of ions co-moving with the local fluid flow $\vec u$, respectively. The first term models the Fickian diffusion, which is the movement of ions from regions of high concentration to regions of low concentration due to Brownian motion. $z_\pm = \pm 1$ denotes the valency of the $\text{K}^+$ and $\text{Cl}^-$ ions, respectively.

The flow velocity $\vec u$ is given by Stokes' equations
\begin{equation}
\begin{aligned}
\label{eq:ekin_hydrodynamics}
\eta \bm{\nabla^{2}} \vec{u} &= \nabla p - \textstyle\sum_\pm \vec j_\pm^\text{diff}/\mu_\pm ,\\
\nabla \cdot \vec{u} &= 0 ,
\end{aligned}
\end{equation}
which model momentum transport due to viscous friction in an incompressible Newtonian fluid, in our case water with a viscosity $\eta = \SI{8.94e-4}{\pascal\s}$. Here $p$ denotes the hydrostatic pressure. Due to the small size (\si{\micro\m}) and low flow velocities (\si{\milli\m\per\s}) in our microfluidic system, convective momentum transport and compressibility effects are insignificant, as indicated by the small Reynold's number $\text{Re} \approx 10^{-4}$ and Mach number $\text{Ma} \approx 10^{-8}$. In addition, the fluid flow relaxes on a time scale much faster than the inhomogeneities in the ionic concentrations or the motion of DNA, which makes Stokes' equations~\eqref{eq:ekin_hydrodynamics} an excellent model for this kind of flow. The coupling force $\vec f = \textstyle\sum_\pm \vec j_\pm^\text{diff}/\mu_\pm$ chosen here represents the drag force exerted by the ions onto the fluid and is responsible for driving EOF. Using this fluid coupling over formulations comprising only electrostatic forces leads to improved stability in numerical algorithms, while producing the same solutions for the flow velocity.~\cite{rempfer_limiting_2016}


\subsection{Finite-Element Model}
\label{sec:fem}

Numerically solving the electrokinetic equations \mbox{(Eqs.~\ref{eq:ekin_electrostatics}-\ref{eq:ekin_hydrodynamics})} is challenging because of the large discrepancy between the length scale of the double layer (\si{\nano\m}) and the system geometry (\si{\centi\m}). We employ the finite-element method (FEM) using a highly adaptive mesh in combination with a carefully crafted representative geometry to limit the computational effort to a level that allows us to investigate a vast parameter space.

\Cref{fig:system}b depicts this representative geometry, whose rotational symmetry we take advantage of to reduce the computational cost. The geometry consists of a conical microcapillary with length $L = \SI{20}{\micro\m}$, inner opening angle $\alpha = \ang{5}$, and orifice radius $r_{\text{m}} = \SI{0.5}{\micro\m}$. The capillary extrudes from a circular barrier of radius $R = 2.5 L$ and width $W = 0.15 L$, separating two hemispherical reservoirs. The capillary carries a surface charge $\sigma_{\text{p}} = \SI{-0.02}{\coulomb\m^{-2}}$, 
which fades out smoothly into the uncharged barrier, and which we later vary. Both the capillary and barrier are impermeable to ions and act as no-slip boundaries for the fluid. The relative permittivity of the solution $\varepsilon_{\text{r}} = 78.54$ is homogeneous and chosen to be that of water, while the pore and barrier permittivities $\varepsilon_{\text{p}} = 4.2$ match that of borosilicate glass. A voltage $U = \pm \SI{1}{\V}$ and a pressure difference $P = \SI{-250}{\pascal}$ is applied over the capillary via the hemispherical edges of the reservoirs, which are also used to impose the bulk salt concentration $c_s$ for both ionic species. Hydrodynamic exchange of momentum with the reservoir is prevented by imposing zero normal stress as boundary condition for the flow at these boundaries.
\vspace{-3mm}
\subsection{Validation and Limitations}

We perform a detailed analysis of the influence of geometric parameters, including the length of the capillary, the smoothing of the capillary tip, the barrier thickness, the shape of the electrodes, the surface charge smoothing, and the reservoir size. For a length of $L = \SI{20}{\micro\m}$, there is a $\SI{2}{\percent}$ deviation from the result for an infinitely long pore (established via finite-size-scaling). Thus our simulations are representative of the experiment described in \Cref{sec:experiment}, where the pore is centimeters long.\\
Using the resulting solutions for the flow field $\vec u$ and the electric field $\vec E$, we determine a field of trajectories for DNA molecules of any given electrophoretic mobility by integrating their equation of motion~\eqref{eq:point_particle}. We validate this approach with simulations of a charged sphere with a diameter of \SI{10}{\nano\m} explicitly represented as a spherical boundary on the symmetry axis in the finite-element simulation and find excellent agreement.\\\indent
While the approach described in Section \ref{sec:dna_motion} to \ref{sec:fem} allows us to evaluate trajectory fields for a wide range of system parameters, we do so using an important simplification. Since the \mbox{\textlambda-phage} DNA is not represented in the FEM simulation for the electric field $\vec E$ and fluid flow $\vec u$, we assume there to be no electrostatic or hydrodynamic interactions between different \mbox{\textlambda-phage} DNA molecules. Furthermore, \cref{eq:point_particle} models \mbox{\textlambda-phage} DNA molecules as point particles, which neglects excluded volume interactions and interactions of different parts of the same molecule. The lack of the repulsive electrostatic and excluded volume interactions leads to minor differences between the experimentally observed DNA distributions and their simulated counterparts as shown in \Cref{sec:results}.


\section{Results}
\label{sec:results}

\subsection{Experimental Observations}
\label{sec:exp_results}

\begin{figure}
\includegraphics[width=\columnwidth]{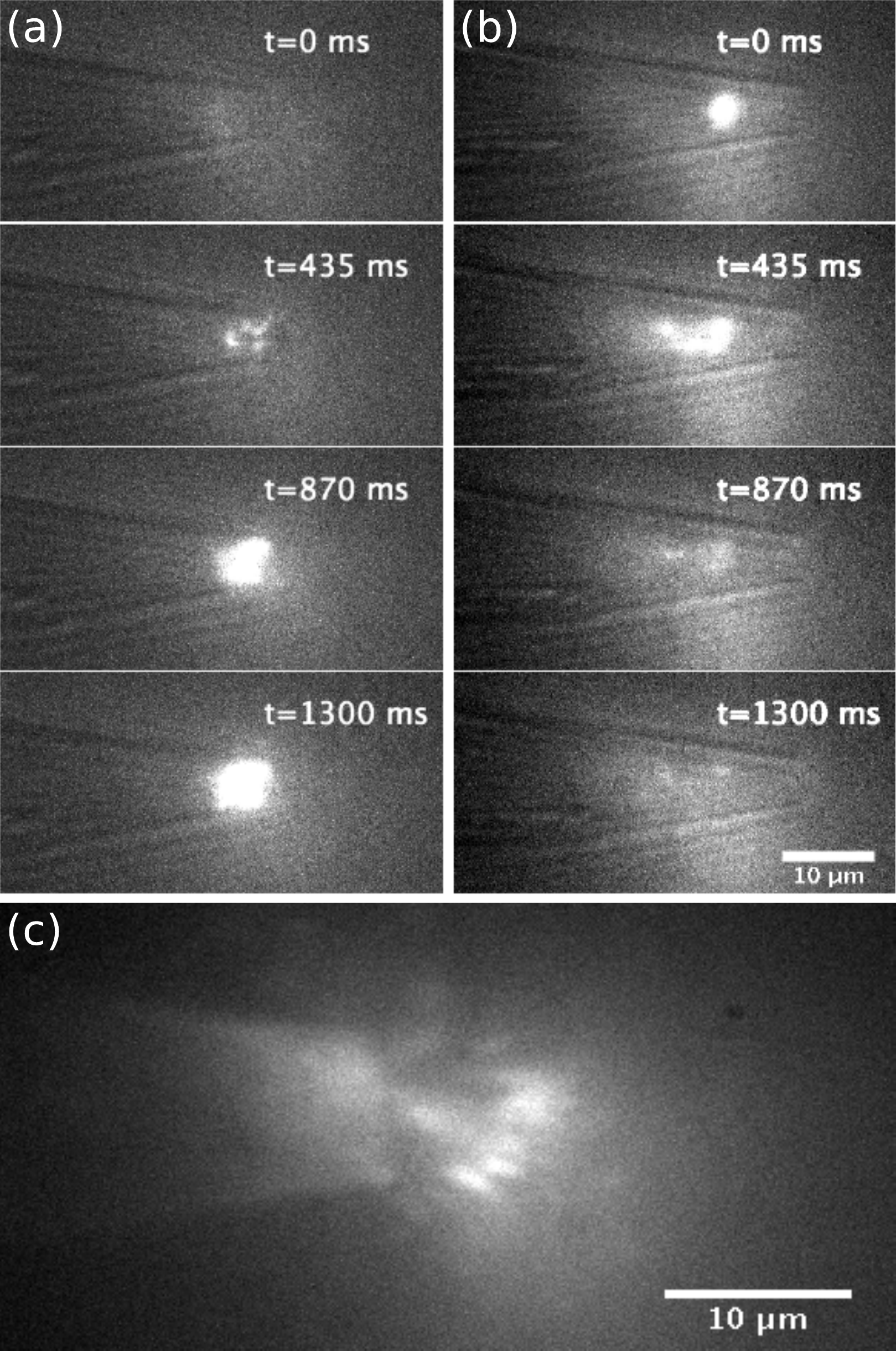}
\caption{Fluorescence microscopy images of the tapered end of the capillary in a sample cell as described in \Cref{sec:experiment} and \ref{sec:exp_results}. Lighter colors represent regions of high fluorescent intensity (high DNA concentration), while dark colors represent regions of low fluorescent intensity (low concentration of DNA).
\textbf{(a)} Capillary tip at different times after applying a positive voltage U = \SI{500}{\milli\volt} (electric field directed out of the capillary / to the right). DNA accumulates in the capillary tip as indicated by the growing fluorescent intensity in this region.
\textbf{(b)} Capillary tip at different times after reversing the voltage (electric field directed into the capillary / to the left). Under these conditions the previously accumulated DNA disperses. Part of it moves into the capillary (to the left), while another part leaves through the capillary orifice.
\textbf{(c)} Pressure-driven translocation of \mbox{\textlambda-phage} DNA through a capillary with a larger orifice radius of \SI{8}{\micro\m}.
}
\label{fig:exp_images}
\end{figure}

We use sample cells as described in \Cref{sec:experiment} with a salt concentration of $\SI{10}{\milli\Molar}$ KCl and apply a positive voltage of \SI{500}{\milli\volt} between the capillary and the reservoir containing the \mbox{\textlambda-phage} DNA (electric field and EOF directed out of the capillary tip). At a distance of several \si{\micro\m} from the capillary orifice, the motion of \mbox{\textlambda-phage} DNA  is diffusion dominated. Within a radius of a few \si{\micro\m}, negatively charged \mbox{\textlambda-phage} DNA moves to the capillary orifice by means of electrophoresis. Instead of translocating through the orifice and moving into the capillary, large amounts of \mbox{\textlambda-phage} DNA accumulate in the capillary tip over a time of \SI{1300}{\milli\s} and remain there, as indicated by the large fluorescent intensity observed at the capillary tip in the images in Fig.\,\ref{fig:exp_images}a.\\\indent
\begin{figure*}
\begin{minipage}[t]{\textwidth}
\includegraphics[width=8.64468cm]{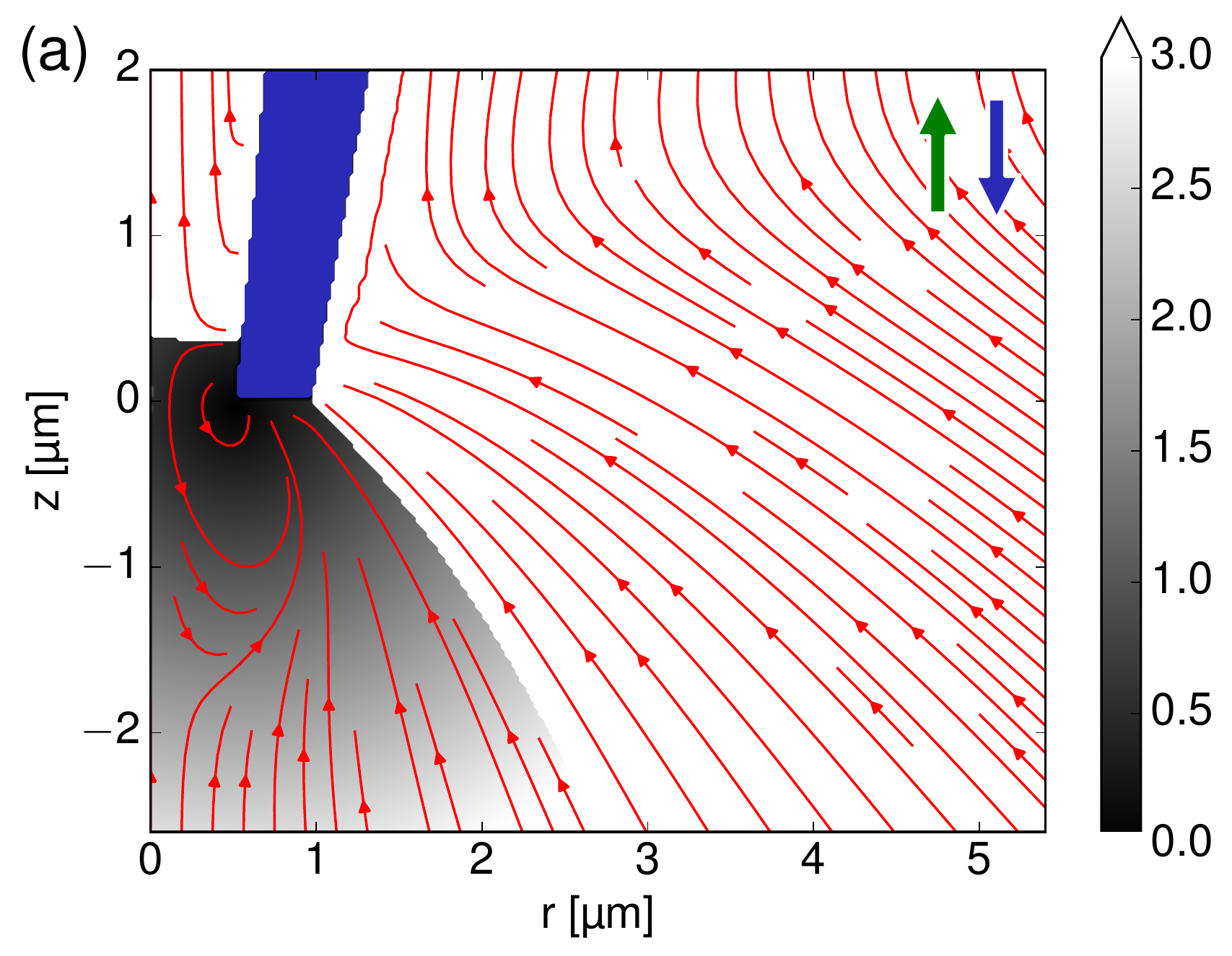}\hfill\includegraphics[width=8.64468cm]{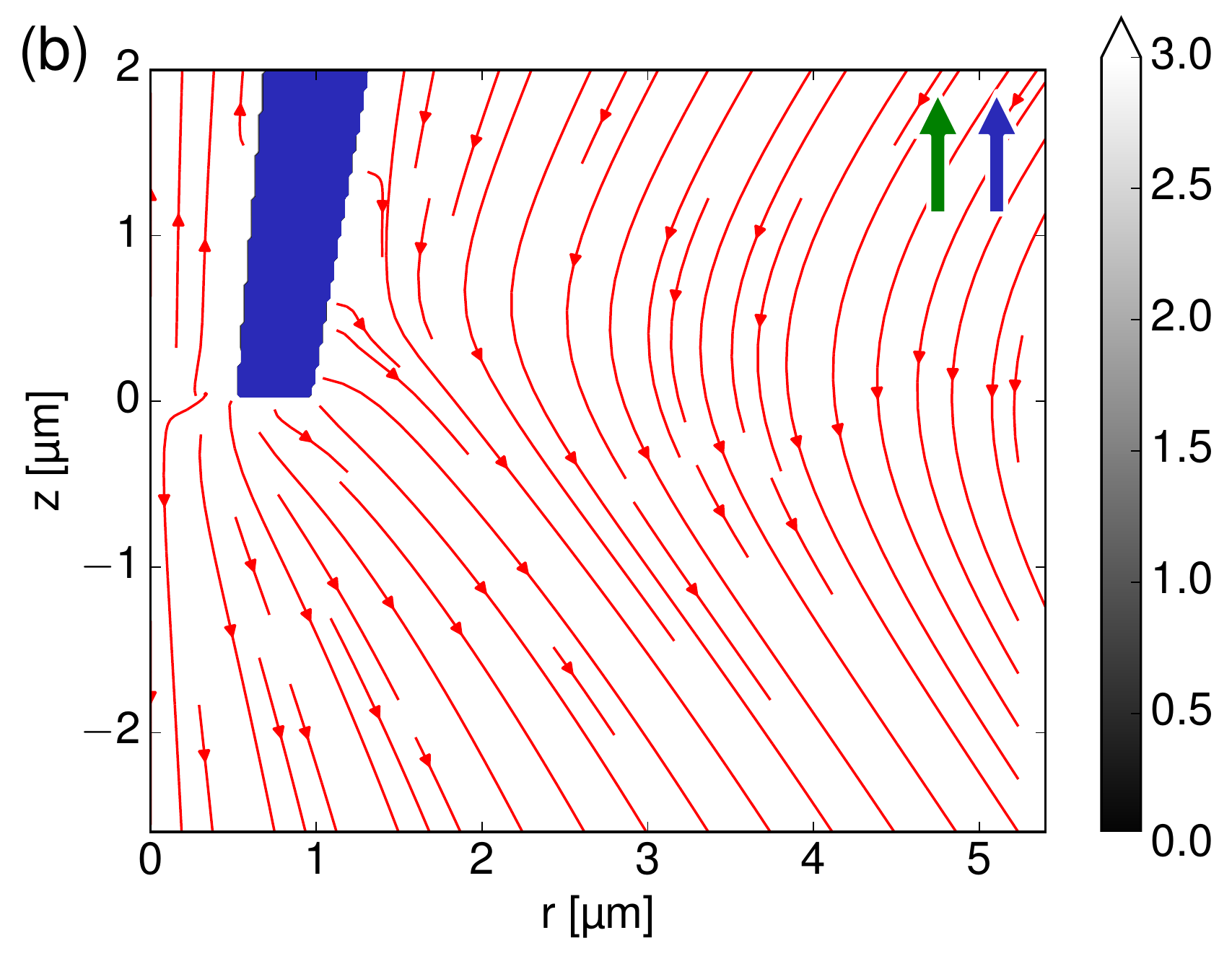}
\caption{Trajectories of analyte particles (red streamlines) at the tip of the capillary (blue) obtained through FEM simulations and the equation of motion~\eqref{eq:point_particle}. The green arrow denotes the direction of the pressure-driven flow in the capillary, while the blue arrow denotes the direction of the electric field and the EOF in the capillary. The system parameters other than the applied voltage correspond to the optimized parameter set given in \Cref{sec:sim_optimized}. The jagged edges are an artifact of the visualization; the simulation's resolution is much higher. \textbf{(a)} Recirculating trapping region at the capillary tip formed for an applied voltage $U = \SI{1}{\volt}$. The brightness of the shaded background (grayscale) is proportional to the distance a molecule travels from the given point (color scale in \si{\micro\m}). Consequently, the black/dark-gray denotes the trapping region, light gray the volume from which analyte molecules move into the trapping region, and points from which analyte molecules escape are shown in white. \textbf{(b)} System for an applied voltage $U = \SI{-1}{\volt}$. The white background and the red trajectories show that analyte molecules move up into the capillary and down into the reservoir without becoming trapped.}
\label{fig:sim_trap_split}
\end{minipage}
\end{figure*}
In the next step, we immediately invert the applied voltage, switching from an electric field directed out of the capillary to one directed into the capillary, as shown in the first image ($t = \SI{0}{\milli\s}$) of Fig.\,\ref{fig:exp_images}b. As the images in the right column at progressively later times show, the accumulated DNA breaks up: part of it exits through the capillary orifice back into the reservoir (to the right of the image); part of it moves further into the capillary (to the left of the image).\\\indent
In addition to controlling the motion and accumulation of \mbox{\textlambda-phage} DNA using an electric field, we can also control its behavior using pressure-driven flow. To create a pressure bias between the two reservoirs, we vary the filling height of the two reservoirs. This creates pressure-driven flow through the capillary superimposed with the EOF induced by the electric field.\\\indent
As discussed in detail in section \Cref{sec:simulations}, EOF produces a plug flow profile in the capillary, whose flow magnitude is independent of the capillary diameter. On the other hand, the flow magnitude of the pressure-driven Poiseuille flow profile scales quadratically with the capillary diameter. We therefore expect pressure-driven flow to dominate the transport of DNA molecules for capillaries of large diameter, and a combination of EOF and electrophoresis to dominate for capillaries of small diameter. Capillaries with intermediate diameters are the most interesting for potential trapping applications, since they should allow us to use both voltage and pressure differences to control the DNA motion.\\\indent
We again trap \mbox{\textlambda-phage} DNA at the capillary tip by applying a positive voltage to the working electrode. We then observe how the behavior of the \mbox{\textlambda-phage} DNA changes when we vary the direction and magnitude of a superimposed pressure-driven flow. We find that for small orifice diameters of up to \SI{4}{\micro\m}, the DNA remains trapped at the capillary tip, but the shape of the DNA accumulation can be heavily influenced by the pressure-driven flow. We conclude that for these orifice diameters, the magnitude of pressure-driven flow and EOF are comparable. \Cref{fig:exp_images}c shows the experiment carried out using capillaries with a larger orifice diameters of \SI{8}{\micro\m}. As expected for these large orifice diameters, we observe pressure-driven translocation of \mbox{\textlambda-phage} DNA, independently of the applied voltage.

\subsection{Numerical Modeling of DNA Trapping}
\label{sec:sim_optimized}

Using the experimental parameters for the applied voltage, pressure bias, salt concentration, and capillary orifice radius as a starting point, we carried out a number of simulations, as described in \Cref{sec:simulations}, to identify a set of system parameters and a particle mobility that result in a pronounced trapping effect. This optimization results in the system depicted in Fig.\,\ref{fig:sim_trap_split}a with parameters of $U = \SI{1}{\volt}$, $P = \SI{-250}{\pascal}$, $c_s = \SI{1}{\milli\Molar}$, and $\sigma_\text{p} = \SI{-0.02}{\coulomb\m^{-2}}$, and an electrophoretic mobility $\mu = \SI{-9.70e-4}{\square\centi\meter\per\volt\per\second}$ for the analyte molecules, somewhat higher than that of \mbox{\textlambda-phage} DNA with $\mu_\text{DNA} = \SI{-6.23e-4}{\square\centi\meter\per\volt\per\second}$ at \SI{1}{\milli\Molar} salt concentration.~\cite{hoagland_capillary_1999}  All of these parameters are experimentally relevant.

\Cref{fig:sim_trap_split}a shows a small subvolume of the simulation domain at the tip of the capillary for the system with the aforementioned optimized parameters. The blue area represents the tapered capillary tip, the red streamlines depict the trajectories of a small charged particle with the above mobility $\mu$. The trajectories were obtained using the equation of motion~\eqref{eq:point_particle}, with the electric field and flow profile from the FEM simulation as input. The green and blue arrows denote the direction of the pressure-driven flow and the EOF on the inside of the capillary, respectively.

Note that with the low salt concentrations and complex geometry used here, the interplay between the ionic distributions, electrostatics, and hydrodynamics can lead to unexpected polarization effects and even reversal of the EOF in the reservoir.~\cite{laohakunakorn_electroosmotic_2015-1}

The shape of the trajectories clearly shows transport of analyte molecules from the reservoir bulk to the capillary tip and a region directly at and in front of the capillary orifice, where they continuously recirculate. This effect directly corresponds to strong accumulation of highly concentrated DNA observed in the experiment (Fig.\,\ref{fig:exp_images}a). 

Interestingly, the complicated interplay of EOF, pressure-driven flow, and electrophoresis does not lead to a simple stalling of analyte molecules at the capillary tip. Previous investigations based on one-dimensional transition state theory as carried out by Hoogerheide \emph{et al.}~\cite{hoogerheide_pressurevoltage_2014}\ for solid state nanopores can therefore not be applied to this capillary based system. White \emph{et al.}~\cite{lan_nanoparticle_2011, german_controlling_2013, lan_effect_2014} have carried out studies using FEM based models similar to the one presented here. They investigated the nanopore translocation of a colloidal particle explicitly represented in the FEM simulation. They relied on the cylindrical symmetry of the system and could therefore not study the off-axis dynamics of the colloidal particle. \citet{tsutsui_particle_2016} investigated the off-axis translocation of a colloid through a solid-state nanopore and found significant differences to the on-axis translocation.

The gray-shaded background depicts the distance an analyte particle travels from a given point. The motion of particles in the recirculation region is limited to distances smaller than this trapping region. Consequently, the trapping region is shaded in black/dark-gray, according to its diameter of \SIrange{1}{2}{\micro\m}. The region shaded in lighter gray marks the subvolume from which analyte particles are transported into the trapping region. This capture region extends over \SI{33}{\micro\m} in the negative z-direction, all the way to the boundary of the simulation domain. Even at that boundary, the analyte molecules' velocities due to advection and electrophoresis range from \SIrange{1}{2.5}{\micro\m\,\s^{-1}}, significantly above the average thermal velocity of \mbox{\textlambda-phage} DNA of \SI{0.57}{\micro\m\,\s^{-1}}. Analyte molecules originating in the white sub-volume do not become trapped: those molecules initially located more than \SI{0.4}{\micro\m} deep inside the capillary tip move up into the capillary, while particles located in the white region on the outside of the capillary move up along the outer capillary surface.\\\indent
The size and position of the simulated trapping region in the left panel of Fig.\,\ref{fig:sim_trap_split}a differs from that of the experimentally observed \mbox{\textlambda-phage} DNA accumulation. The reasons for this are two-fold. Firstly, the optimized trapping parameters determined here do differ somewhat from the experimentally used parameters: the salt concentration is lower (\SI{1}{\milli\Molar} instead of \SI{10}{\milli\Molar}), the applied voltage higher (\SI{1}{\volt} instead of \SI{500}{\milli\volt}), and the pore orifice diameter smaller (\SI{1}{\micro\m} instead of \SI{4}{\micro\m}). Such a change in the parameters to achieve trapping is expected, due to the relative physical simplicity of our model. Secondly, the simulation neglects the repulsive electrostatic and excluded volume interactions between different DNA molecules as discussed in \Cref{sec:simulations}. The lack of repulsive interactions and the smaller orifice radius explain why the size of the trapping region in the simulations is smaller than that of the DNA accumulation observed in the experiment (Fig.\,\ref{fig:exp_images}a).

\Cref{fig:sim_trap_split}b depicts the system with reversed voltage ($U = \SI{-1}{\volt}$) and otherwise identical parameters. The trajectories split up into two populations from where they were previously recirculating. One part of the molecules, which were previously trapped, now moves up into the capillary, while the other part disperses back into the reservoir. This situation also directly corresponds to the experimental observations as depicted in Fig.\,\ref{fig:exp_images}b. We therefore conclude that our simulation model accurately captures the relevant physics of the experimental system observed and described in~\Cref{sec:exp_results}.

\subsection{Tuning the Trap for Specific Molecules}
\label{sec:sim_tuning}

\begin{figure*}[t]
\centering
\begin{tabular}{@{}cc@{}}
\includegraphics[width=\columnwidth]{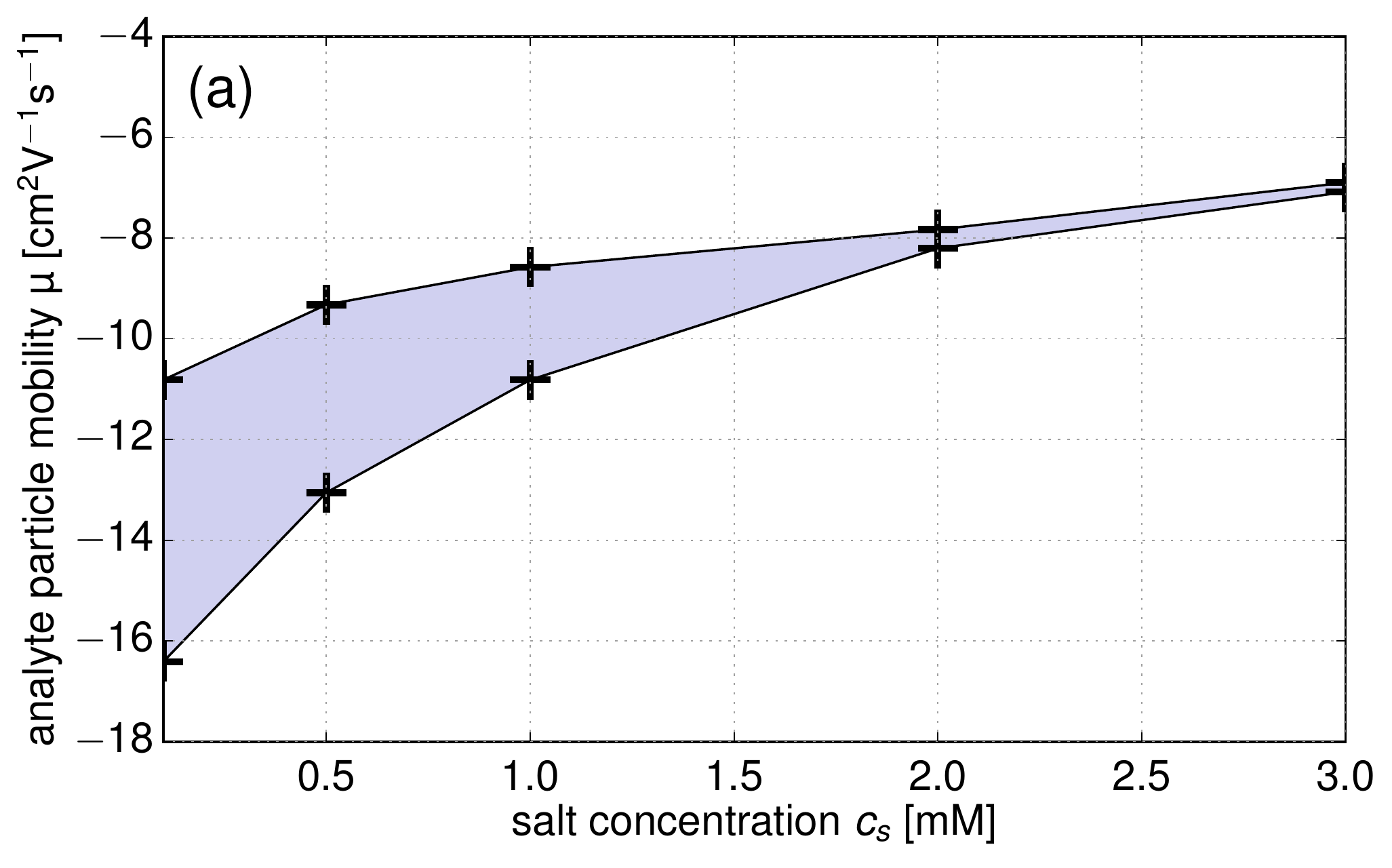} &
\includegraphics[width=\columnwidth]{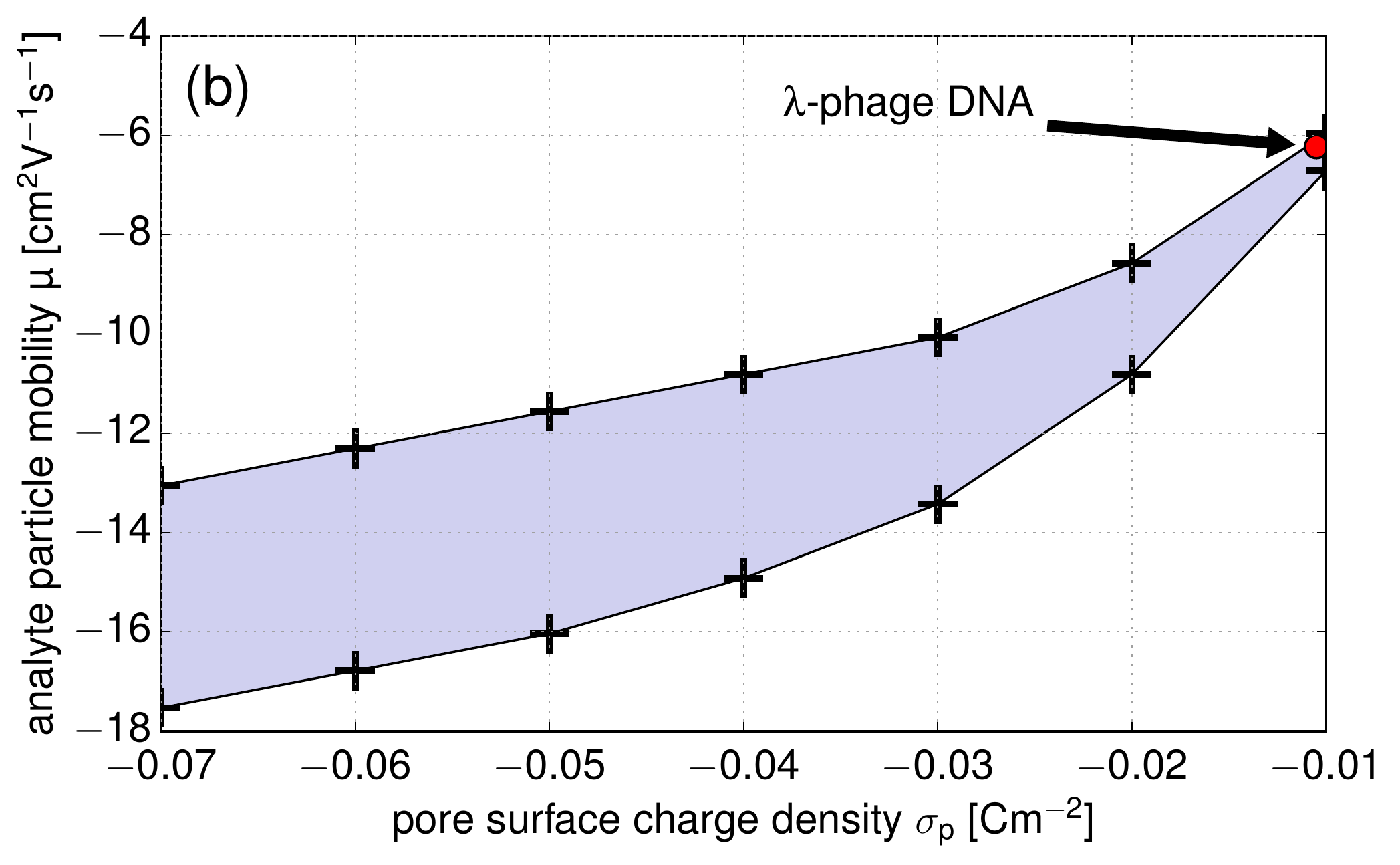} \\
\includegraphics[width=\columnwidth]{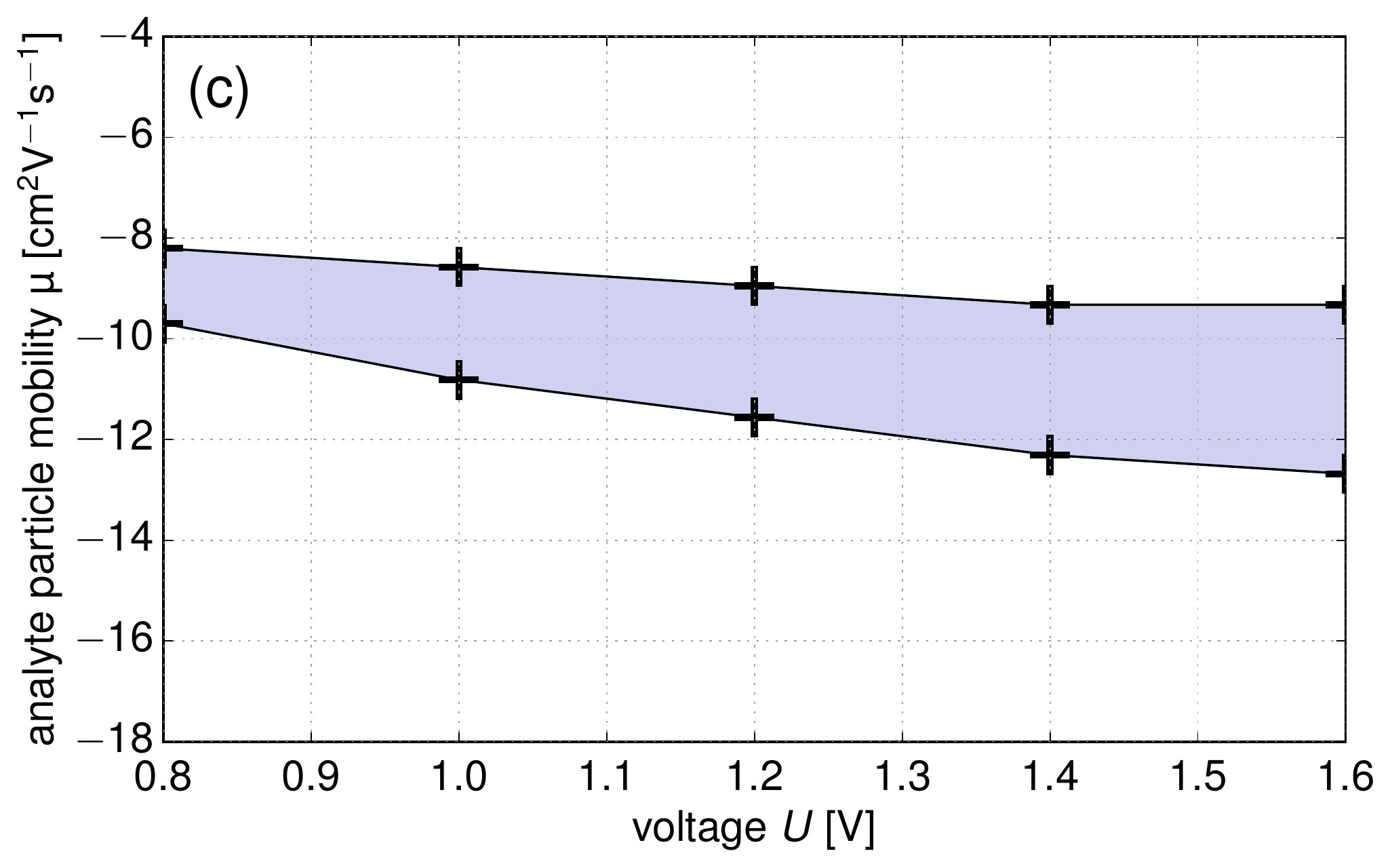} &
\includegraphics[width=\columnwidth]{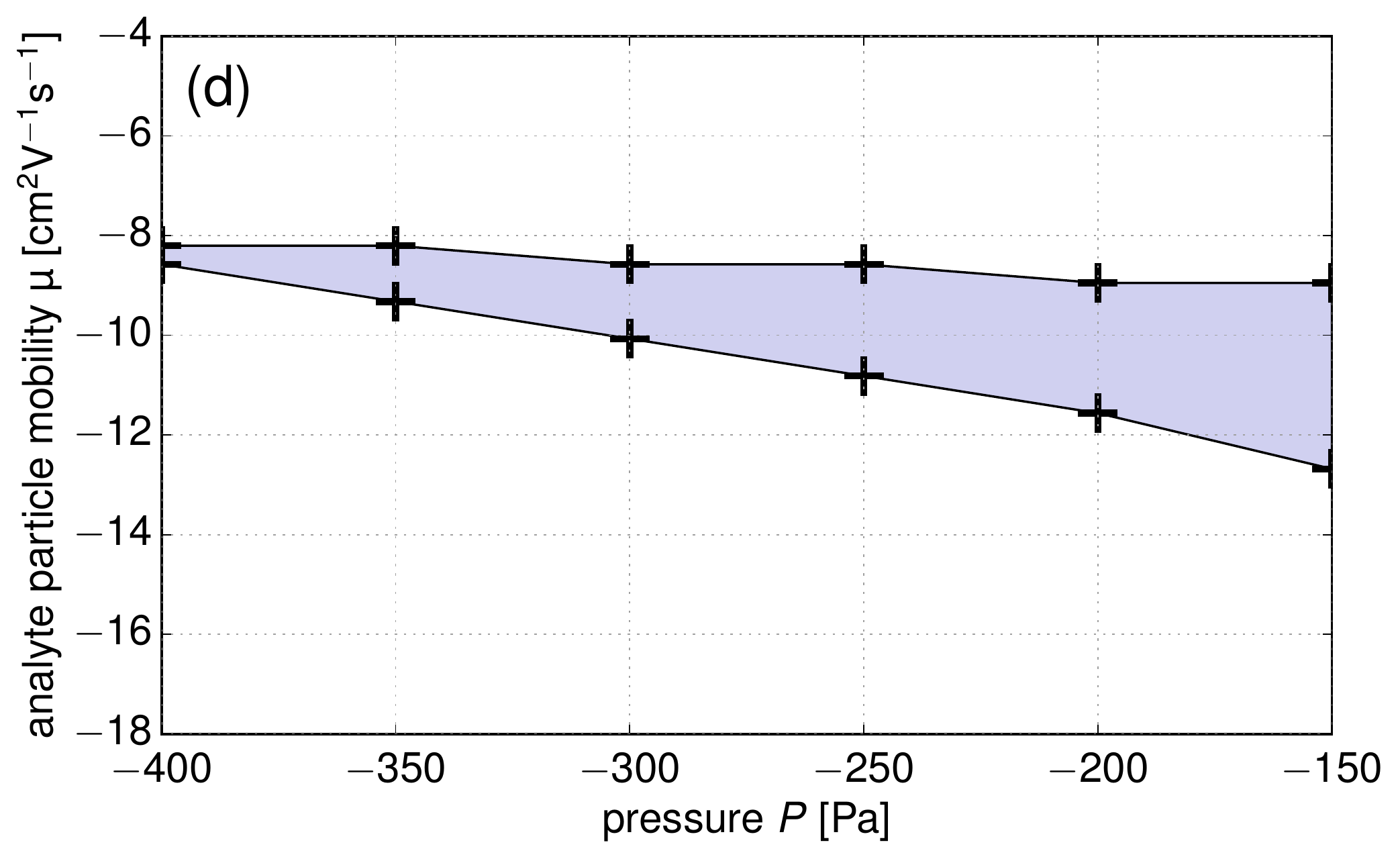}
\end{tabular}
\caption{Range of electrophoretic mobility of captured particles at the capillary tip. Each of the panels varies one of the system parameters for the salt concentration $c_s$ (a), the capillary surface charge density $\sigma_\text{p}$ (b), the applied voltage $U$ (c), and the applied pressure bias $P$ (d). The remaining three parameters are taken from the optimized parameters in \Cref{sec:simulations}: $c_s = \SI{1}{\milli\Molar}$, $\sigma_\text{p} = \SI{-0.2}{e\per\nano\m^2}$, $U = \SI{1}{\volt}$, and $P = \SI{-250}{\pascal}$. Simulations were carried out for the parameter values denoted by the black pluses. The blue shaded area marks the range of mobility of trapped particles. The red dot in the panel (b) indicates the mobility of \mbox{\textlambda-phage} DNA and shows that the trap can be tuned to capture \mbox{\textlambda-phage} DNA.}
\label{fig:sim_selectivity}
\end{figure*}

Having identified a set of parameters for this microcapillary based DNA trap, we now perform a numerical, exploratory study to determine whether this system can be tuned to trap a wide range of very specific molecules. We restrict ourselves to a numerical study here, as this is the most suited to consider a large parameter space and affords the most control over the variation of external parameters. While the fabrication method for the capillaries is relatively inexpensive and facile, performing a multitude of experiments on different types of molecules and varying external parameters is time consuming and costly, thus going well beyond the scope of the current work. \\
To determine the generality of capillary trapping, we varied the system parameters for the voltage, the pressure difference, the salt concentration, and the capillary surface charge, starting from the optimized parameters identified in \Cref{sec:sim_optimized}. For each of these systems, we then investigated the resulting trajectory field for its suitability as a trap, depending on the electrophoretic mobility of analyte molecules. This procedure yields upper and lower bounds for the trapping mobility of analyte molecules as a function of the varied system parameters. We refer to lower and higher mobilities in terms of their absolute values, that is, higher mobilities are more negative, while lower mobilities are mobilities closer to zero.\\
As Fig.\,\ref{fig:sim_selectivity} demonstrates, only particles with negative mobilities become trapped for all of the investigated parameters. Due to symmetry, positive analyte molecules can be trapped by applying the opposite pressure and voltage to a device of the same geometry but with opposite capillary surface charge.

Fig.\,\ref{fig:sim_selectivity}a shows how the mobilities of trapped analyte particles vary with the salt concentration. We find that at low salt concentrations of \SI{0.1}{\milli\Molar}, a relatively wide range of mobilities between \SI{-1.08e-3}{\square\centi\meter\per\volt\per\s} and \SI{-1.64e-3}{\square\centi\meter\per\volt\per\s} leads to trapping. Moderately increasing the salt concentration to \SI{3}{\milli\Molar} gradually shifts the interval for the mobilities of trapped particles to lower values and continuously increases the specificity of the trap. At a salt concentration of \SI{3}{\milli\Molar}, only particles with mobilities in the range from \SI{-6.90e-4}{\square\centi\meter\per\volt\per\s} to \SI{-7.09e-4}{\square\centi\meter\per\volt\per\s} become trapped. Experimentally varying the salt concentration within this range from \SI{0.1}{\milli\Molar} to \SI{3}{\milli\Molar} is very feasible.\\\indent
Fig.\,\ref{fig:sim_selectivity}b shows the range of mobilities of trapped particles as a function of the capillary surface charge density. Varying the pore surface charge density from \SIrange{-0.07}{-0.01}{\coulomb\m^{-2}} allows us to tune the mobilities of trapped particles to very similar bounds as before when varying the salt concentration from \SI{0.1}{\milli\Molar} to \SI{3}{\milli\Molar}. At a surface charge concentration of  \SI{-0.07}{\coulomb\m^{-2}}, analyte particles with mobilities in the range from \SI{-1.31e-3}{\square\centi\meter\per\volt\per\s} to \SI{-1.75e-3}{\square\centi\meter\per\volt\per\s} become trapped. With decreasing capillary surface charge density (in absolutes), this interval gradually moves to lower mobilities and the trap becomes more specific. At a capillary surface charge density of \SI{-0.01}{\coulomb\per\m^2}, only particles with mobilities in the range from \SI{-5.97e-4}{\square\centi\meter\per\volt\per\s} to \SI{-6.71e-4}{\square\centi\meter\per\volt\per\s} become trapped. This range includes the mobility of \mbox{\textlambda-phage} DNA $\mu_\text{DNA} = \SI{-6.23e-4}{\square\centi\meter\per\volt\per\second}$ at \SI{1}{\milli\Molar} salt concentration.~\cite{hoagland_capillary_1999}
Varying the spatial distribution of the capillary surface charge density experimentally is non-trivial. However, surface charge can be controlled changing the pH, or by using a wide range of available polyelectrolyte coatings, or even coatings with lipid membranes. All of these surface coatings can suppress or enhance EOF, leading to an effective surface charge that can be tuned over a wide range.\\\indent
Fig.\,\ref{fig:sim_selectivity}c depicts the dependency of the trapped particle's mobility as a function of the applied voltage. We vary the applied voltage from \SIrange{0.8}{1.6}{\volt} and find that in contrast with the other parameters, the voltage in this range does not change the specificity of the trap drastically. At \SI{0.8}{\volt}, particles with mobilities between \SI{-8.21e-4}{\square\centi\meter\per\volt\per\s} and \SI{-9.70e-4}{\square\centi\meter\per\volt\per\s} become trapped. With increasing voltage, this interval moves to higher mobilities. At \SI{1.6}{\volt}, particles with mobilities between \SI{-9.32e-4}{\square\centi\meter\per\volt\per\s} and \SI{-1.27e-3}{\square\centi\meter\per\volt\per\s} become trapped. Electrolysis of water happens for voltages exceeding \SI{1.23}{\volt}, negatively impacting the experiment due to pH shifts and bubble formation at the electrodes causing current instabilities.~\cite{atkins_elements_2013}\\\indent
Finally, Fig.\,\ref{fig:sim_selectivity}d shows how the applied pressure difference influences the mobilities of trapped particles. The lower limit for the mobility of trapped particles is only weakly affected by varying the pressure difference in the range from \SIrange{-400}{-150}{\pascal} and varies from \SI{-8.21e-4}{\square\centi\meter\per\volt\per\s} to \SI{-8.58e-4}{\square\centi\meter\per\volt\per\s}. The upper limit of mobilities leading to trapping, on the other hand, is significantly affected by that change in pressure and changes from \SI{-8.58e-4}{\square\centi\meter\per\volt\per\s} at a pressure of \SI{-400}{\pascal} to \SI{-1.27e-3}{\square\centi\meter\per\volt\per\s} at a pressure of \SI{-150}{\pascal}. These pressure differences can easily be achieved experimentally by closing the reservoirs and connecting them to a microfluidic pressure regulator.\\\indent
Our results in Fig.\,\ref{fig:sim_selectivity} therefore suggest that varying the salt concentration or the pore surface charge density within experimentally plausible values offers the greatest specificity for trapping analyte molecules. Pressure variations in the region of \SIrange{-400}{-350}{\pascal} also offer a reasonable degree of specificity, but varying the voltage has little effect on specificity. This is in a sense fortuitous since voltages exceeding \SI{1.23}{\volt} cause the undesirable electrolysis of water. Our glass nanopore system therefore shows strong promise as an experimental useful and cheap system for trapping charged macromolecules with a high degree of control.

\section{Conclusions}
\label{sec:conclusion}

We have shown experimentally that inexpensive glass microcapillaries can be used to accumulate \mbox{\textlambda-phage} DNA at their tip through a combination of EOF, pressure-driven flow, and electrophoresis. We further showed that this accumulated DNA can be transported into and through the capillary simply by reversing the applied voltage.\\\indent
Having demonstrated this effect experimentally, we then investigated the ability of this system to accumulate specific macromolecules using finite-element computer simulations based on the electrokinetic equations. We identified a set of parameters for the salt concentration, surface charge density of the capillary, applied voltage, pressure difference, and capillary orifice diameter that exhibit a strong trapping effect for particles with experimentally relevant electrophoretic mobilities.\\\indent
Also in line with the experiments, the trap discontinues when the electric field reverses and the accumulated DNA splits into two lumps: one moving into the capillary and one moving away from it into the bulk.\\\indent
In our simulations, we go beyond the on-symmetry-axis approximation that is typically made in the literature to-date~\cite{hoogerheide_pressurevoltage_2014, lan_nanoparticle_2011, german_controlling_2013, lan_effect_2014} to study the trapping of nanoparticles and other analytes. We find that this off-axis approach is crucial to understanding the way particles trap. Specifically, our traps are not static structures, instead, a region in which the DNA moves in a recirculating pattern close to the tip forms when the right combination of external pressure and electric-field is applied. Thus, the on-axis result can lead to false impressions of the physics of trapping at nano- and microcapillary tips.\\\indent
Our results demonstrate that a trap using glass microcapillaries can be finely tuned to concentrate very specific macromolecules from solution. This, together with the cheap and facile fabrication of our system, makes glass microcapillaries show a great deal of promise for the use as analytic devices.\\\indent
Our results will guide further experiments and we believe that they may ultimately serve as the blueprint for the cheap, simple filtering and pre-concentrating of analyte solutions in microfluidic lab-on-a-chip devices.

\section*{Acknowledgments}

GR and CH thank the DFG for funding through the SFB716/TPC.5. NL acknowledges support from the George and Lillian Schiff Foundation, and Trinity College, Cambridge. JdG gratefully acknowledges financial support by an NWO Rubicon Grant (\#680501210).

\bibliography{references}

\end{document}